\documentclass[12pt]{article}
\usepackage{hyperref}
\usepackage{graphicx}
\usepackage{color}
\usepackage{amsmath,amssymb} 

\definecolor{myblue}{rgb}{0.14,0.11,0.49}
\definecolor{myred}{rgb}{0.74,0.22,0.15}
\definecolor{mygreen}{rgb}{0.05,0.52,0.42}
\definecolor{myyellow}{rgb}{0.96,0.92,0.13}
\definecolor{myorange}{rgb}{1,0.61,0.36}
\definecolor{mypurple}{rgb}{0.71,0.02,1}
\definecolor{noir}{gray}{0.} 

\newcommand{\Couleur}[1]{\textcolor{myblue}{#1}} 
\definecolor{htc}{rgb}{1,1,1} 

\newcommand{\Mat}[1]{{{\boldsymbol{#1}}}}
\newcommand{\abs}[1]{\left\vert#1\right\vert}
\def\be{\begin{equation}}
\def\ee{\end{equation}}
\def\bea{\begin{equationarray}}
\def\eea{\end{equationarray}}
\def\bc{\begin{center}}
\def\ec{\end{center}}
\def\bi{\begin{itemize}}
\def\ei{\end{itemize}}
\def\bs{\begin{slide}}
\def\es{\end{slide}}
\def\dd{\mathrm{d}}
\def\iC{\mathrm{i}}

\begin{document}
\title{Towards testing a dark matter candidate that emerges from the scalar ether theory}
\author{
Mayeul Arminjon\\
\small\it Univ. Grenoble Alpes, CNRS, Grenoble INP
, 3SR, F-38000 Grenoble, France.\\
\small E-mail: mayeul.arminjon@3sr-grenoble.fr
}
\date{}
\maketitle

\begin{abstract}
\noindent 
According to a scalar theory of gravity with a preferred frame, electromagnetism in the presence of a gravitational field implies that there is an additional energy tensor, which might contribute to dark matter. The expression of this tensor is determined by a mere scalar $p$, that depends on the EM field and (for a weak field) on the Newtonian gravitational field. We briefly recall why this tensor arises and how the EM field in a galaxy can be calculated. The data fields that enter the PDE for the scalar field $p$ oscillate very quickly in space and time, as does the EM field. This prevents integration of that PDE at the relevant galactic scale. Therefore, a homogenization of that PDE has to be operated. We discuss in some detail three possible ways of applying the homogenization theory to that PDE: time, space, or spacetime homogenization. The second and third ways may lead to feasible, albeit heavy calculations.
\end{abstract}
\section{Introduction}

Although an unseen, exotic form of matter seems to make up the dominant part in the mass of galaxies and clusters of galaxies, the problem of identifying what this ``dark matter" could be made of has been and remains one of the big enigmas in contemporary physics. The reason for the present work is that it turns out \cite{A57} that, according to an alternative, relativistic theory of gravity with a preferred frame: In the presence of both a gravitational field and an electromagnetic (EM) field, there must indeed appear some exotic ``interaction energy" \Couleur{$E_\mathrm{inter}$}, which should be distributed in space, and be gravitationally active. This has been found while searching \cite{A54,A56,A57} for a consistent formulation, in that theory of gravity, of the Maxwell equations in the presence of gravitation --- the aim was thus not to invent a new candidate for dark matter. \\

That energy \Couleur{$E_\mathrm{inter}$} could nevertheless contribute to the dark matter. It depends on the EM field \Couleur{$({\bf E},{\bf B})$} and the gravity field. The theory of gravity that we refer to, based on a scalar field only, is called ``scalar ether theory" or \underline{{\bf SET}}; see Ref. \cite{A35}. To check if \Couleur{$E_\mathrm{inter}$} might indeed build a ``dark halo" around a galaxy, we should be able to compute the Interstellar Radiation Field (hereafter \underline{{\bf ISRF}}) as a solution of the Maxwell equations --- because the equations that govern \Couleur{$E_\mathrm{inter}$} \cite{A57} contain the Maxwell field \Couleur{$({\bf E},{\bf B})$} and its first-order derivatives. However, the existing models for the ISRF focus on radiation transfer (e.g. \cite{Draine1978, Mathis-et-al1983, Chi-Wolfendale1991, Gordon-et-al2001, Robitaille2011, Popescu-et-al2017}) and do not produce an EM field \Couleur{$({\bf E},{\bf B})$}, even less one that would be an exact solution of the Maxwell equations. Hence, we built from scratch a model that does produce an exact solution of the Maxwell equations \cite{A61}. That model is based on axial symmetry (of the galaxy and the ISRF) as a relevant approximation. It turns out that there exists an explicit representation for all axisymmetric solutions of the source-free Maxwell equations \cite{A60}. The model makes predictions for the Spectral Energy Density (SED) of the ISRF, that are close to those of the existing models \cite{A62}. Except for one thing: the new model predicts extremely high values of the SED on the galaxy's axis \cite{A63}.\\

Our current work, of which this paper is a progress report, is as follows: Using the axisymmetric model of the ISRF in a galaxy as an exact Maxwell EM field, just mentioned, we seek to compute the interaction energy density field \Couleur{$E_\mathrm{inter}$} in the weak gravitational field of a galaxy. Beyond the EM field, \Couleur{$E_\mathrm{inter}$} depends also on the Newtonian potential \Couleur{$U$}, more precisely on \Couleur{$\partial_t (\nabla U)$}. The time derivative \Couleur{$\partial_t $} is taken in the preferred frame (``ether") \Couleur{$\mathcal{E}$}. Hence the field \Couleur{$E_\mathrm{inter}$} will depend on the velocity \Couleur{${\bf V}$} of the center of the galaxy w.r.t. \Couleur{$\mathcal{E}$}.\\

Before exposing our present work, in the next two sections we shall briefly summarize the foregoing steps \cite{A57,A60,A61}.


\section{Interaction tensor in SET}

The equations of electrodynamics on general relativity (GR) simply rewrite those of special relativity (SR) by using the ``comma goes to semicolon" rule: \Couleur{$\quad _{,\,\nu} \ \rightarrow \ _{;\,\nu}$}, i.e.:  partial derivatives are replaced by covariant derivatives. That is not possible in SET, for the Dynamical Equation for the energy(-momentum-stress) tensor \Couleur{$\Mat{T}$} is not generally  \Couleur{$ T^{\lambda \nu }_{\ \, ;\nu}=0$} (which rewrites \Couleur{$T^{\lambda \nu }_{\ \, ,\nu}=0$} valid in SR). In SET, the first group of the Maxwell equations is unchanged. The second group \underline{was} obtained by applying the Dynamical Equation of SET to a charged medium in the presence of Lorentz force, assuming that (as is the case in GR): 
\be\label{Ass1}
\Couleur{\mathrm{\ Total\ energy\ tensor}\ \Mat{T} = \Mat{T}_\mathrm{charged\ medium} + \Mat{T}_\mathrm{field}}. \\
\ee
The additivity (\ref{Ass1}) leads to a form of Maxwell's second group in SET \cite{A54,B39}. But that form of Maxwell's second group in SET predicts charge production/destruction at untenable rates, and hence has to be discarded \cite{A56}. The additivity assumption (\ref{Ass1}) is contingent and may be abandoned. \hypertarget{Intro-T_Inter}{It means introducing an} ``interaction" energy tensor \Couleur{$\Mat{T}_\mathrm{inter}$} such that 
\be\label{Tinter}
\Couleur{\Mat{T}_{(\mathrm{total})} = \Mat{T}_\mathrm{charged\ medium} + \Mat{T}_\mathrm{field}\ \underline{+ \Mat{T}_\mathrm{inter}}\,}. 
\ee
In the work \cite{A57}, it has been found that the assumption that
 \Couleur{$\Mat{T}_\mathrm{inter}$} should be Lorentz-invariant in the situation of special relativity (SR), i.e. when the ``physical" spacetime metric \Couleur{$\Mat{\gamma}$} is Minkowski's metric \Couleur{$\Mat{\gamma}^0$} (\Couleur{$\gamma^0 _{\mu \nu }=\eta _{\mu \nu }$} in Cartesian coordinates), leads unambiguously to the following definition in the general case:
\be\label{T_inter_mixed}
\Couleur{T^\mu_{\mathrm{inter}\ \ \nu }:=  p\,\delta ^\mu _\nu}, \qquad \mathrm{or}\quad \Couleur{(T_\mathrm{inter})_{\mu  \nu }:=  p\,\gamma  _{\mu \nu}}.
\ee
(Note that this is generally-covariant.) It has also been shown that the additional equation made necessary by the additional unknown: the scalar field \Couleur{$p$}, can then be consistently imposed to be the equation for {\it charge conservation}. In that way, the electrodynamics of SET is a closed system of PDE's that satisfies charge conservation. The ``interaction energy"
\be
\Couleur{E_\mathrm{inter}:=T^{0 0}_\mathrm{inter} = p\gamma ^{0 0}} 
\ee
may be regarded as (macroscopic) ``dark matter", because (i) it is not localized inside usual matter: we have \Couleur{$ p\ne 0$} at a generic point; (ii) it is gravitationally active: \Couleur{$T_\mathrm{inter}^{0 0}\,$} contributes to \Couleur{$T^{0 0}$} which is the source of the gravitational field in SET; (iii) it is not usual matter (e.g. no velocity can be defined for the medium having \Couleur{$T^\mu_{\mathrm{inter}\ \ \nu }$} as energy tensor, since, when considered in coordinates that are Cartesian at the event considered, the tensor (\ref{T_inter_mixed}) is invariant under local Lorentz transformations).

Moreover, the scalar PDE that determines the first approximation of the scalar field \Couleur{$p$} in a given general EM field and in a given weak gravitational field with Newtonian potential \Couleur{$U$} has been obtained. Its initial form is (Eq. (69) in Ref. \cite{A57}):
\footnote{\ 
Equation (\ref{Eq for p}) is valid in coordinates \Couleur{$x^\mu$} that are Cartesian for the flat spacetime metric \Couleur{$\Mat{\gamma }^0$} of SET and adapted to the preferred frame. Thus the time coordinate is \Couleur{$x^0 = cT$} with \Couleur{$T$} the preferred time of the theory, and the spatial coordinates are Cartesian for the Euclidean spatial metric \Couleur{$\Mat{g }^0$} of SET \cite{A57}. The l.h.s. of (\ref{Eq for p}) is invariant under a Lorentz transformation of \Couleur{$\Mat{\gamma }^0$}, but the r.h.s. is so only if that transformation is internal to the preferred frame.
}
\be\label{Eq for p}
\Couleur{\left (-G^{\mu \nu }\, p_{,\nu}\right )_{,\mu}= -f},
\ee
where \Couleur{$G^{\mu \nu }$} (noted \Couleur{$G^{\ \, \mu \nu }_1$} in Ref. \cite{A57}) are the components of an antisymmetric spacetime tensor \Couleur{$\Mat{G}$}: the inverse tensor of the EM field tensor made with the components of \Couleur{$({\bf E},{\bf B})$} --- the EM field of the first approximation, that obeys the flat-spacetime Maxwell equations. In addition, in Eq. (\ref{Eq for p}), we have
\be
\Couleur{f: = c^{-3} \left( e^i \partial _T U \right)_{,i }\left (1+O\left(c^{-2}\right)\right)},
\ee
in which \Couleur{$e^i\ (i=1,2,3)$} are the components  of a spatial vector \Couleur{${\bf e}$} also made with the components of \Couleur{$({\bf E},{\bf B})$}. Using the antisymmetry of \Couleur{$G^{\mu \nu }$}, Eq. (\ref{Eq for p}) can be rewritten \cite{A57} as:
\begin{equation}\label{dp/dt}
\Couleur{\left(\frac{\dd p}{\dd t} \right )_{\bf u} := \frac{\partial p}{\partial  t} + {\bf u.}\nabla p = S},
\end{equation}
in which the source field \Couleur{$S$} and the spatial vector field \Couleur{${\bf u}$} are given, both depending on \Couleur{$({\bf E},{\bf B})$}, and with the source \Couleur{$S$} depending also on \Couleur{$\nabla \partial _T U$}. Thus, in principle, \Couleur{$p$} obtains by integrating \Couleur{$S$} along the curves \Couleur{$\frac{\dd {\bf x}}{\dd t} = {\bf u}(t,{\bf x})$} \cite{A57}. \\



\section{Maxwell model of the ISRF}

To check if \Couleur{$E_\mathrm{inter}$} might build a ``dark halo", we must have the Interstellar Radiation Field in a galaxy (ISRF) as a Maxwell field. In Ref. \cite{A61}, a model has been proposed to calculate the ISRF as an exact solution of the Maxwell equations. Note that this is of astrophysical interest, independently of the problem of checking what has just been mentioned. See e.g. Ref. \cite{A63}, Sect. 2, for a short presentation of the model. Here we shall provide only the main elements of that model, as follows:\\

(i) We assume {\it axial symmetry}. This is a relevant approximation for many galaxies. The \Couleur{$z$} axis is taken as the symmetry axis.\\

(ii) We consider the {\it source-free} Maxwell equations. Indeed, we want to describe the ISRF at a galactic scale, not in the stars (which are the primary source of the ISRF) or in their neighborhood. \\

(iii) We consider a finite set of frequencies \Couleur{$(\omega _j)\ (j=1,...,N_\omega )$}, thus a finite number of time-harmonic source-free Maxwell fields. The number of frequencies of a simulation, \Couleur{$N_\omega $}, determines how accurately the continuous spectrum is represented for that simulation.\\

(iv) We use a theorem \cite{A60} that expresses any time-harmonic axisymmetric source-free Maxwell field as the sum of two such fields, obtained from two scalar potentials \Couleur{$A_z$} and \Couleur{$A'_z$}. Moreover, in the relevant ``totally-propagating" case, each of these potentials depends only on a ``spectrum" function of one variable, \Couleur{$S=S(k)$} with \Couleur{$k$} the axial wave number. In addition, for a galaxy, the logic of the model (points (v) and (vi) below) leads one to assume that \Couleur{$A_z=A'_z$} \cite{A63}, thus one spectrum function \Couleur{$S_j$} for each frequency \Couleur{$\omega _j\ (j=1,...,N_\omega)$}. \\

(v) An axisymmetric galaxy is schematized as a finite set \Couleur{$\{{\bf x}_i\,;\  i=1,...i_\mathrm{max} \} $} of point-like ``stars", given by their cylindrical coordinates \Couleur{$\rho ,\phi ,z$}, which are drawn with specific probability laws ensuring axisymmetry and representativity.\\

(vi) The model is fixed by the spectrum functions \Couleur{$S_j \ (j=1,...,N_\omega)$}, approximated by their discretized values \Couleur{$S_{n j} \ (n=1,...,N)$}. For each frequency \Couleur{$\omega _j$}, to determine the values \Couleur{$S_{n j} \ (n=1,...,N)$}, we do a least-squares fitting of a sum of spherical potentials:
\be
 \Couleur{\Sigma_j:=\sum_{i=1} ^{i_\mathrm{max}} \varphi  _{{\bf x}_i\,\omega_j}}.
\ee
The potential \Couleur{$\varphi  _{{\bf x}_i\,\omega_j}$} is the outgoing spherical solution of the wave equation with frequency \Couleur{$\omega_j$}, emanating from the star at point \Couleur{${\bf x}_i$} . \\

\section{Calculation of \Couleur{${\bf u}$} and \Couleur{$S$}}

In order to integrate Eq. (\ref{dp/dt}), we have to be able to compute the values of the vector field \Couleur{${\bf u}$} and the source \Couleur{$S$} at any given event \Couleur{$(t,{\bf x})$}. The explicit expressions of \Couleur{${\bf u}$} and \Couleur{$S$} were given in Ref. \cite{A57}. Setting  \Couleur{$\Pi : = {\bf E.B}$}, we get immediately from Eqs. (73)-(77) there, that
\footnote{\label{EB ne 0}\
We must assume that $ \Couleur{{\bf E.B} \ne 0}$ \cite{A54}. It is not valid for the simple and most known solutions of the Maxwell equations: purely electric or purely magnetic fields, usual plane waves, dipole field \cite{A56}. However, it is generally true, because a real EM field is a combination of simple solutions and, if one adds two solutions \Couleur{$({\bf E}_1 , {\bf B}_1)$},  \Couleur{$({\bf E}_2 , {\bf B}_2)$} of the standard Maxwell equations such that \Couleur{${\bf E}_1 . {\bf B}_1=0$} and \Couleur{${\bf E}_2 . {\bf B}_2=0$}, then \Couleur{$({\bf E}_1 + {\bf E}_2,{\bf B}_1 + {\bf B}_2)$} is also a solution, but 
\be
 \Couleur{({\bf E}_1 + {\bf E}_2).({\bf B}_1 + {\bf B}_2)= {\bf E}_1. {\bf B}_2 + {\bf E}_2.{\bf B}_1 },
\ee
which generally is not zero. In short, \Couleur{${\bf E.B}$} depends non-linearly on the field \Couleur{$({\bf E}, {\bf B})$}.
}
\be\label{u=}
\Couleur{{\bf u}= \frac{{\bf E}\wedge \nabla \Pi -   ({\partial _t}\Pi) {\bf B}}{{\bf B.}\nabla \Pi}},
\ee
\be\label{S=}
\Couleur{S =  \frac{\mathrm{div} \left ( {\bf e}\, \partial_t U \right )} {c^3 {\bf B.}\nabla \frac{1}{\Pi}}},
\ee
where the spatial (3-)vector \Couleur{${\bf e}$} (Eq. (68) in Ref. \cite{A57}) is easily shown to be
\be
\Couleur{{\bf e} = \frac{1}{c \mu_0} \left({\bf E} + \frac{c^2 {\bf B}^2-{\bf E}^2}{2\Pi}{\bf B} \right )}
\ee
(here \Couleur{$\mu_0$} is the vacuum permeability). Since \Couleur{$A'_{j z}=A_{j z}$} for the model EM field, it follows that here \Couleur{$c^2 {\bf B}^2-{\bf E}^2 =0$}, see Eq. (19) in Ref. \cite{A63}. Thus \Couleur{${\bf e} = {\bf E}/(c \mu_0) $}. With the free Maxwell equations, we have \Couleur{$\mathrm{div} \,{\bf E} =0$}, so the numerator of (\ref{S=})  is \Couleur{$ {\bf E.}\nabla(\partial_t U)/(c \mu_0)$}.\\

The calculation of \Couleur{${\bf u}$} was numerically implemented, and tried with the model of the Milky Way, already investigated \cite{A61,A62,A63}. As could be expected in view of Eq. (\ref{u=}), the time variation of \Couleur{${\bf u}$} is on the scale of the quasi-period of the EM field, and its space variation is on the scale of its wavelengths, thus both are extremely small as compared with the relevant scales for a galaxy. Therefore, it is numerically unfeasible to integrate on a galactic scale the PDE (\ref{dp/dt}) for the scalar field \Couleur{$p$}.

\section{Homogenization of the PDE for the scalar field \Couleur{$p$}}

We want to obtain information on the field of interaction energy, thus on the scalar field \Couleur{$p$}, at the macroscopic scale of a galaxy. But to do this, what is at our disposal is the PDE (\ref{Eq for p}) or the PDE (\ref{dp/dt}) --- whereas either equation involves fields that vary on a very microscopic scale, e.g. Eq. (\ref{dp/dt}) involves the fields \Couleur{${\bf u}$} and \Couleur{$S$}. This situation is typical of the homogenization theory, which has been developed mostly for periodic media --- see e.g. Refs. \cite{BLP978,SanchezPalencia1980,Caillerie2012}. The aim of this kind of theory is to get ``homogenized" PDE's allowing one to describe the medium at the macroscopic scale. Note that for Eq. (\ref{dp/dt}) the ``medium" is, essentially, characterized by the pair of given heterogeneous fields \Couleur{$({\bf u},S)$}. For Eq. (\ref{Eq for p}), the ``medium" is characterized by the pair of given heterogeneous fields \Couleur{$(\Mat{G},f)$}. Actually, for a real galaxy, that ``medium" is quasi-periodic rather than periodic, but the structure of the homogenized equations is the same for a quasi-periodic medium as for a truly periodic medium \cite{Caillerie2012,Auriault2011}. The most important difference is that, in the first case, the elementary cell and/or the data fields within the cell, vary with the macroscopic position. (See in Subsects. \ref{Time-hom}--\ref{Spacetime-hom} below for the relevant definitions of the elementary cell.)  We note also that the homogenization theory has often been applied to the EM field, e.g. \cite{SanchezPalencia1980,Barbatis-Stratis2003}, thus obtaining homogenized Maxwell equations, that apply to some specific situation --- in the case of Refs. \cite{SanchezPalencia1980,Barbatis-Stratis2003}, an anisotropic linear medium with permittivity and permeability tensors that vary with the position at a very small scale (and in a periodic way). However, these works do not seem to be relevant to the problem of homogenizing the PDE's (\ref{Eq for p}) or (\ref{dp/dt}). Two obvious points are: (i) a galaxy (thus a set of stars in the first place) cannot be regarded as a medium of the kind just mentioned. (ii) What has to be homogenized here, is the PDE for the scalar field \Couleur{$p$}, not the Maxwell equations for the ISRF. Anyway, the ISRF in a galaxy does oscillate at the extremely small time and space scales mentioned above.\\

The homogenization technique considers two variables related by a small parameter $\epsilon$: a ``slow" variable, say $s$, which is the one that browses the medium at the macroscopic scale as is sought for, and the ``quick" variable, $s/\epsilon$, of which an $O(1)$ variation will browse the period of the medium, which is very small as evaluated in terms of the slow variable. Several different frames can thus be envisaged for the homogenization of the PDE's (\ref{Eq for p}) or (\ref{dp/dt}), depending on which set of variables is considered for the homogenization: the time variable, the space variables, or the spacetime variables. 

\subsection{Time homogenization of Eq. (\ref{dp/dt})}\label{Time-hom}

We will first illustrate the technique in the case of the time variable, thus we will outline the application to Eq. (\ref{dp/dt}) of the {\it time homogenization} technique as proposed (for the mechanics of heterogeneous materials) by Guennouni \cite{Guennouni1988}.
The idea of the method is to consider two ``separated" time scales: a ``quick" time \Couleur{$\tau$} and a ``slow" time \Couleur{$t$}, with \Couleur{$\tau = t/T$}, where \Couleur{$T\ll 1$} is the period of the quick variation (here that of the EM radiation field), when expressed in terms of the ``slow" time variable \Couleur{$t$}. Thus \Couleur{$T\ll t$} --- here the galactic time scale: \Couleur{$t\simeq r/c$} with \Couleur{$r$} a galactic distance, hence the ratio \Couleur{$t/T \simeq r/\lambda$} has a huge value  \Couleur{$\simeq 10^{25}$}: we have here a very good separation of scales. The elementary cell is one interval of periodicity, thus \Couleur{$[0,T]$} in terms of the slow variable \Couleur{$t$}, and \Couleur{$[0,1]$} in terms of the quick variable \Couleur{$\tau =t/T$}. Formally, one assumes that the given fields \Couleur{${\bf u}$} and \Couleur{$S$}, as well as the boundary values for \Couleur{$p$} (which shall not be precised), have the form 
\be\label{u and S periodic}
\Couleur{{\bf u}={\bf u}({\bf x},\,t,\,\tau) = \lambda (t,\tau) {\bf u}^*({\bf x})},\quad \Couleur{S=S({\bf x},\,t,\,\tau) = \mu (t,\tau) S^*({\bf x})}, ...
\ee
where \Couleur{$\lambda $} and \Couleur{$\mu $} are \Couleur{$\tau$}-periodic of period \Couleur{$1$}.\\

Setting 
\be
\Couleur{{\bf u}^T({\bf x},t) := {\bf u}\left ({\bf x},\,t,\,\frac{t}{T}\right )}, ..., 
\ee
one has a boundary value problem \Couleur{$\pi^T$} for \Couleur{$p$}, depending smoothly on \Couleur{$T$}. Its solution field \Couleur{$p^T({\bf x},t)$} defines a field depending on two time variables:
\be
\Couleur{p\left ({\bf x},\,t,\, \tau \right ) := p^T({\bf x},t)}\quad \mathrm{with\ } \Couleur{T = \frac{t}{\tau}}.
\ee
The total time derivative is 
\be\label{total timeder}
\Couleur{\frac{\dd p^T}{\dd t} = \frac{\partial }{\partial t} \left (p \left({\bf x},t,\frac{t}{T} \right )\right)= \frac{\partial p}{\partial t} + \frac{1}{T} \frac{\partial p}{\partial \tau}}.
\ee
One states an asymptotic expansion as \Couleur{$T\rightarrow 0$} for the unknown field:
\be\label{p expansion}
\Couleur{p^T({\bf x},t) = p_0({\bf x},t,\tau) T^0 + p_1({\bf x},t,\tau) T + O(T^2)} \quad \left(\Couleur{\tau = \frac{t}{T}}\right), 
\ee
where \Couleur{$p_0$} and \Couleur{$p_1$} are \Couleur{$\tau$}-periodic of period \Couleur{$1$}. Note that, since \Couleur{$\lambda $} and \Couleur{$\mu $} in Eq. (\ref{u and S periodic}) are assumed \Couleur{$\tau$}-periodic, the given fields \Couleur{${\bf u}$} and \Couleur{$S$} are ``already developed": Eq. (\ref{u and S periodic}) provides their expansions, those expansions having only the \Couleur{$T^0$} term. Using (\ref{total timeder}) and (\ref{p expansion}) in (\ref{dp/dt}) and identifying powers, we get
\be\label{expanded eqs}
\mathrm{order\ }\Couleur{T^{-1}:\ \frac{\partial p_0}{\partial \tau} = 0\ i.e.\ p_0=p_0({\bf x},t)},\quad \mathrm{order\ }\Couleur{T^{0}:\ \frac{\partial p_0}{\partial t}+\frac{\partial p_1}{\partial \tau} + {\bf u.}\nabla p_0=S}.
\ee
Averaging the last equation over the period \Couleur{$T$}, according to
\be
\Couleur{\bar{f}({\bf x},t):=\int_0 ^1 f({\bf x},t,\tau) \dd \tau },
\ee
yields 
\be\label{average eq} 
\Couleur{\frac{\partial p_0}{\partial t}+ \bar{{\bf u}}{\bf .}\nabla p_0=\bar{S}},
\ee
the sought-for time-averaged equation. Note that, from (\ref{expanded eqs})$_1$, it follows that \Couleur{$p_0=\bar{p}$}. Thus, the time-homogenized equation (\ref{average eq}) differs from the ``microscopic" starting equation (\ref{dp/dt}) merely by the fact that time-averaged fields are substituted for the fields entering (\ref{dp/dt}), both for the given fields \Couleur{${\bf u}$} and \Couleur{$S$} and for the unknown field \Couleur{$p$}. This was not {\it a priori} obvious: a naive averaging of Eq. (\ref{dp/dt}) would leave us with \Couleur{$\overline{{\bf u}.\nabla p}$} in the place of \Couleur{$\bar{{\bf u}}{\bf .}\nabla p_0 \equiv \bar{{\bf u}}{\bf .}\nabla \bar{p}$}.\\

Thus, the application of the time homogenization technique to Eq. (\ref{dp/dt}) is quite straightforward, at least formally. (We did not investigate the convergence as \Couleur{$T\rightarrow 0$}.) However, the data fields that are obtained after the time-averaging, \Couleur{$\bar{{\bf u}}$} and \Couleur{$\bar{S}$}, have still the very rapid {\it spatial} variation, at the scale of the typical wavelength. As a result: as it is the case for the starting equation (\ref{dp/dt}), again it is numerically unfeasible to integrate Eq. (\ref{average eq}) on a galactic scale. This has been checked numerically, after having implemented the calculations of the time averages of \Couleur{${\bf u}$} and \Couleur{$S$}. We give some indications about the latter calculations below.

\subsubsection{Time-averaging  \Couleur{${\bf u}$} and \Couleur{$S$}}

The Maxwell model of the ISRF provides each of the six components of \Couleur{${\bf E}$} and \Couleur{${\bf B}$} in the form
\be\label{F(t)}
\Couleur{F^{(q)}(t,{\bf x}) = {\mathcal Re} \left ( \sum _{j=1} ^{N_\omega } C^{(q)}_j({\bf x}) e^{-\iC \omega _j t} \right )\qquad (q=1,...,6)}.
\ee
Then, \Couleur{$\Pi : = {\bf E.B}$} expands on the \Couleur{$e^{-\iC (\omega _j + \omega _k)  t}$} 's and \Couleur{$e^{-\iC (\omega _j - \omega _k)    t}$} 's. Hence, \Couleur{${\bf E}\wedge \nabla \Pi$},  \Couleur{$({\partial _t}\Pi) {\bf B}$}, and \Couleur{${\bf B.}\nabla \Pi $}, which enter (\ref{u=}) and (\ref{S=}), expand on the \Couleur{$e^{-\iC \Omega _\kappa   t}$} 's and \Couleur{$e^{-\iC \Psi _\kappa     t}$} 's, with \Couleur{$\kappa =(j,k,m),\quad \Omega _\kappa= \omega _j + \omega _k + \omega _m,\quad \Psi_\kappa = \omega _j + \omega _k - \omega _m$}. 
Those three fields time-average to zero, because \Couleur{$\omega _j + \omega _k - \omega _m $} does not vanish, not any more than (of course) \Couleur{$\omega _j + \omega _k + \omega _m $} . But equations (\ref{u=}) and (\ref{S=}) for \Couleur{${\bf u}$} and \Couleur{$S$} involve {\it ratios} of these fields. Thus we have to compute the time average of a ratio of two trigonometric polynomials: 
\footnote{\ 
The indices in Eq. (\ref{Q(t)}) browse in general a different set as compared with the set \Couleur{$\{1,...,N_\omega \}$} of the frequencies of the EM field.
}
\be
\Couleur{Q(t)} =  \Couleur{\frac{{\mathcal Re} \left(\sum_j C_j e^{-\iC \omega _j t}\right) }{{\mathcal Re} \left(\sum_k D_k e^{-\iC \omega _k t}\right) }} = \Couleur{\frac{\sum_j C_j e^{-\iC \omega _j t} + C^\star_j e^{\iC \omega _j t}}{\sum_k D_k e^{-\iC \omega _k t} + D^\star_k e^{\iC \omega _k t}}} \nonumber
\ee
\be\label{Q(t)}
 = \Couleur{\sum_j \frac{1}{\sum_k \frac{D_k}{C_j} e^{-\iC (\omega _k-\omega_j) t} + \frac{D^\star_k}{C_j} e^{\iC (\omega _k+\omega_j) t}} + \frac{1}{\sum_k \frac{D_k}{C^\star_j} e^{-\iC (\omega _k+\omega_j) t} + \frac{D^\star_k}{C^\star_j} e^{\iC (\omega _k-\omega_j) t}}}.
\ee
We compute that, to an often quite good approximation, 
\be
\Couleur{\left\langle \frac{1}{a+\sum_\kappa b_\kappa e^{-\iC \Omega _\kappa t}}  \right\rangle \simeq \frac{1}{a}}.
\ee
With this approximation, we get
\be
\Couleur{\bar{Q} \simeq \sum_j \frac{1}{\frac{D_j}{C_j}} + \frac{1}{\frac{D^\star_j}{C^\star_j}} = 2 \sum_j {\mathcal Re}\left( \frac{C_j}{D_j} \right)}.
\ee


\subsection{Space homogenization of Eq. (\ref{dp/dt})}\label{Space-hom}

The (space) homogenization of a PDE governing a medium with a fine periodic microstructure has been studied notably in Refs. \cite{BLP978,SanchezPalencia1980}, followed by very numerous works. Here we will outline the space homogenization of Eq. (\ref{dp/dt}) by adapting the work of Caillerie \cite{Caillerie2012}. We assume that the given fields \Couleur{${\bf u}$} and \Couleur{$S$} are defined in the spatial elementary cell \Couleur{$Y$}, a rectangular parallelepiped with sides parallel to the Cartesian axes, as  functions of the ``quick" spatial variable \Couleur{${\bf y}$}, and can be extended to \Couleur{$Y$}-periodic functions of \Couleur{${\bf y}$} defined in the whole space. This means essentially that each of them, along with its derivatives, takes the same value at corresponding points on opposite faces of the parallelepiped \Couleur{$Y$}. As in the time homogenization described above, one considers a boundary value problem for (\ref{dp/dt}) depending on a small parameter \Couleur{$\epsilon$} (in the place of \Couleur{$T$}) --- \Couleur{$\epsilon$} can be considered as the size of the cell (its dimension in the direction of the first coordinate axis, say), that size being expressed in terms of the slow space variable \Couleur{${\bf x}$} such that \Couleur{${\bf y} = {\bf x}/\epsilon$}. To do so one defines
\footnote{\ 
This is somewhat different from Eq. (\ref{u and S periodic}) that was adapted to Eq. (\ref{dp/dt}) from the presentation of the time-homgenization technique in Ref. \cite{Guennouni1988}. However, it is equivalent in practice, because actually we have indeed just one field \Couleur{${\bf u}$}, not a family of fields. Thus Eq. (\ref{u epsilon}) defines such a family \Couleur{$({\bf u}^\epsilon)$} from the data of the unique field \Couleur{${\bf u}$}.
}
\be\label{u epsilon}
\Couleur{{\bf u}^\epsilon({\bf x},t) = {\bf u}\left (\frac{{\bf x}}{\epsilon},t \right )},
\ee
and the like for \Couleur{$S$}. As in Eq. (\ref{p expansion}), one states an asymptotic expansion as \Couleur{$\epsilon \rightarrow 0$} for the solution field:
\be\label{p expansion-space}
\Couleur{p^\epsilon({\bf x},t) = p_0\left({\bf x},{\bf y},t\right) \epsilon^0 + p_1\left({\bf x},{\bf y},t\right) \epsilon + O(\epsilon^2)} , \qquad \Couleur{{\bf y} =\frac{{\bf x}}{\epsilon}},
\ee
where \Couleur{$p_0$} and \Couleur{$p_1$} are \Couleur{$Y$}-periodic in the quick space variable \Couleur{${\bf y}$}. Similarly with Eq. (\ref{total timeder}), the total spatial derivatives are given by
\be\label{total spaceder}
\Couleur{\frac{\dd p^\epsilon }{\dd x^i} = \frac{\partial }{\partial x^i} \left (p \left({\bf x},\frac{{\bf x}}{\epsilon},t \right )\right)= \frac{\partial p}{\partial x^i} + \frac{1}{\epsilon } \frac{\partial p}{\partial y^i}}.
\ee
Then, inserting (\ref{p expansion-space}) in (\ref{dp/dt}) using (\ref{total spaceder}), and identifying powers, we get
\be\label{expanded eqs-space}
\mathrm{order\ }\Couleur{\epsilon^{-1}}:\ \Couleur{\frac{\partial p_0}{\partial y^j} u^j = 0},\quad \mathrm{order\ }\Couleur{\epsilon^{0}:\ \frac{\partial p_0}{\partial t}+\left ( \frac{\partial p_0}{\partial x^j} + \frac{\partial p_1}{\partial y^j} \right ) u^j =S}.
\ee
We note that, in contrast with the time homogenization case (see Eq. (\ref{expanded eqs})$_1$), inserting the expansions (\ref{p expansion-space}) into the PDE (\ref{dp/dt}) does not necessarily imply that \Couleur{$p_0$} is independent of the quick variable \Couleur{${\bf y}$}.
\footnote{\label{Caillerie}\ 
In the case of the stationary heat conduction equation for the temperature \Couleur{$\theta $}, studied in Ref. \cite{Caillerie2012}, the fact that \Couleur{$\theta _0$} is independent of the quick variable is not obtained directly (in contrast with the equation studied in Ref. \cite{Guennouni1988}, as well as for Eq. (\ref{dp/dt}) with the time homogenization). Nevertheless, this fact does follow from the expanded equation of the lowest order, by using the weak form of this equation together with the periodic boundary conditions and the fact that the quadratic form associated with \Couleur{$\Mat{K}$} is coercive (D. Caillerie, private communication). Here the argument valid for the heat equation does not apply and, clearly, Eq. (\ref{expanded eqs-space})$_1$ does not necessarily imply that \Couleur{$\nabla_{\bf y} p_0 = 0$}.
}
However, it is consistent to {\it assume} that \Couleur{$p_0$} is independent of \Couleur{${\bf y}$}: clearly, Eq. (\ref{expanded eqs-space})$_1$ is then automatically satisfied. Another difficulty as compared with the time homogenization case, as well as with the works \cite{Caillerie2012, Guennouni1988}, is the following one. The volume integration, over the elementary cell \Couleur{$Y$}, of the main (zero-order) equation, Eq. (\ref{expanded eqs-space})$_2$, does not in the most general case leave us with a PDE for the averaged fields. Assuming that \Couleur{$p_0$} is independent of \Couleur{${\bf y}$} implies obviously that 
\be
\Couleur{\overline{p_0}({\bf x},t) : = \frac{1}{\abs{Y}} \int_Y p_0({\bf x},{\bf y},t) \,\dd {\bf y} = p_0({\bf x},t) }
\ee
(where \Couleur{$\abs{Y}$} is the volume of \Couleur{$Y$}), so that \Couleur{$P:=p_0 = \overline{p_0}$} is indeed an averaged field. With this, the integration of Eq. (\ref{expanded eqs-space})$_2$ over \Couleur{$Y$} gives us
\be\label{Integ_order_0}
\Couleur{\abs{Y} \left( \frac{\partial P}{\partial t} + \overline{u^j}\,\frac{\partial P}{\partial x^j} \right ) + \int_Y {\bf u.}\nabla_{\bf y} p_1 ({\bf x},{\bf y},t) \,\dd {\bf y} = \abs{Y}\, \overline{S}({\bf x},t)}.
\ee
The integral on the l.h.s. may be rewritten as
\be\label{Integ_u.grad p_1}
\Couleur{\int_Y {\bf u.}\nabla_{\bf y} p_1 ({\bf x},{\bf y},t) \,\dd {\bf y} = \int_{\partial Y} {\bf u.n}\, p_1 \dd \Sigma - \int_Y p_1 \mathrm{div}_{\bf y} {\bf u}  \,\dd {\bf y} = - \int_Y p_1 \,\mathrm{div}_{\bf y} {\bf u}  \,\dd {\bf y}},
\ee
the last equality resulting from the fact that \Couleur{$p_1$} and \Couleur{${\bf u}$} are \Couleur{$Y$}-periodic in \Couleur{${\bf y}$}, so that \Couleur{${\bf u.n}\, p_1$} takes opposite values at corresponding points on opposite faces of \Couleur{$Y$}. However, the integral on the rightmost side of (\ref{Integ_u.grad p_1}) has no apparent reason to vanish in general. If it can be neglected, we can rewrite Eq. (\ref{Integ_order_0}) as
\be\label{Integ_order_0-2}
\Couleur{ \left(\frac{\dd P}{\dd t} \right )_{\overline{{\bf u}}} :=\frac{\partial P}{\partial t} + \overline{u^j}\,\frac{\partial P}{\partial x^j} = \overline{S}({\bf x},t)},
\ee
an equation for the averaged fields. Assuming that the same equation is valid for the relevant case of quasi-periodic fields \Couleur{${\bf u}$} and \Couleur{$S$}, it would then remain to implement the calculation of the spatially-averaged fields \Couleur{$\overline{{\bf u}}$} and \Couleur{$\overline{S}$} and the integration of Eq. (\ref{Integ_order_0-2}).

\subsection{Spacetime homogenization of Eq. (\ref{Eq for p})}\label{Spacetime-hom}

Equation (\ref{Eq for p}) for $p$ has exactly the same form as the stationary heat conduction equation for the temperature \Couleur{$\theta $} (see Eq. (18) in Ref. \cite{Caillerie2012}), except for the fact that here we have a spacetime equation instead of a spatial equation. That is, the tensor \Couleur{$\Mat{G}$} is a spacetime second-order tensor, that plays the role played in Ref. \cite{Caillerie2012} by the spatial second-order tensor of thermal conductivity, \Couleur{$\Mat{K}$}. Therefore, we can easily adapt the work \cite{Caillerie2012}. Again, one considers a boundary value problem for (\ref{Eq for p}) depending on a small parameter \Couleur{$\epsilon$}: \Couleur{$\epsilon$} can be considered as the size of the spacetime elementary cell \Couleur{$\Upsilon$}, that size being expressed in terms of the slow spacetime variable \Couleur{${\bf X} = (t,{\bf x})$}. To do so one defines, similarly to Eq. (21) of Ref. \cite{Caillerie2012}:
\be\label{G epsilon-ST}
\Couleur{\Mat{G}^\epsilon({\bf X}) = \Mat{G}\left (\frac{{\bf X}}{\epsilon} \right )},
\ee
and the like for \Couleur{$f$}. (The latter is a minor difference with Ref. \cite{Caillerie2012}, in which \Couleur{$f$} is considered independent of \Couleur{$\epsilon $}.) 
As in Eqs. (\ref{p expansion}) and (\ref{p expansion-space}), one states an asymptotic expansion as \Couleur{$\epsilon \rightarrow 0$} for the solution field (cf. Eq. (22) of Ref. \cite{Caillerie2012}):
\be\label{p expansion-spacetime}
\Couleur{p^\epsilon({\bf X}) = p_0\left({\bf X},{\bf Y}\right) \epsilon^0 + p_1\left({\bf X},{\bf Y}\right) \epsilon + O(\epsilon^2)} , \qquad \Couleur{{\bf Y} =\frac{{\bf X}}{\epsilon}},
\ee
where \Couleur{$p_0$} and \Couleur{$p_1$} are \Couleur{$\Upsilon$}-periodic in the quick spacetime variable \Couleur{${\bf Y}=(\tau ,{\bf y})$}. A difference with the heat conduction case is that the inverse EM field tensor \Couleur{$\Mat{G}$} is antisymmetric, instead of being symmetric as is the conductivity tensor \Couleur{$\Mat{K}$}. Therefore the quadratic form associated with \Couleur{$\Mat{G}$} is identically zero. This prevents us from using the argument used in the former case (see Footnote \ref{Caillerie}) to show that the first term in the expansion (here $p_0$) is independent of the quick variable, here \Couleur{${\bf Y}$}. As in Subsect. \ref{Space-hom}, we nevertheless still may assume that \Couleur{$p_0 = p_0({\bf X})$}. Then, just the same calculations of Ref. \cite{Caillerie2012} do apply, substituting \Couleur{$\Mat{G}$} for \Couleur{$\Mat{K}$}, \Couleur{$p $} for \Couleur{$\theta $}, and defining, as in Eq. (20) of Ref. \cite{Caillerie2012}:
\be
 \Couleur{{\bf q} : = -\Mat{G}.\nabla_{\bf X} p}. 
\ee
In particular, the following expanded equation is derived (Eq. (25b) in Ref. \cite{Caillerie2012}):
\be\label{expans_eps^0}
\Couleur{\mathrm{div}_{\bf X} {\bf q}_0 + \mathrm{div}_{\bf Y} {\bf q}_1 + f = 0},
\ee
with (Eqs. (26a) and (26b) in Ref. \cite{Caillerie2012}):
\be
\Couleur{{\bf q}_0 = -\Mat{G}.\left (\nabla_{\bf X} p_0 + \nabla_{\bf Y} p_1 \right )},
\ee
\be
\Couleur{{\bf q}_1 = -\Mat{G}.\nabla_{\bf X} p_1 }.
\ee
(The \Couleur{$\nabla_{\bf Y} p_2$} term does not appear, for we stop the expansion (\ref{p expansion-spacetime}) at \Couleur{$p_1$}.) By integrating (\ref{expans_eps^0}) on the spacetime cell \Couleur{$\Upsilon$}, the following homogenized equation is derived (Eq. (27) in Ref. \cite{Caillerie2012}):
\be\label{Eq for p-Hom}
\Couleur{\mathrm{div}_{\bf X}\,\left \langle {\bf q}_0 \right \rangle + \left \langle f \right \rangle = 0},
\ee
with the angular brackets denoting the average over the spacetime cell:
\be
\qquad \Couleur{\left \langle A \right \rangle({\bf X}) :=\int_\Upsilon A({\bf X},{\bf Y}) \dd {\bf Y} / \int_\Upsilon \dd {\bf Y} }.
\ee
Moreover, as with the heat conduction case, the average field \Couleur{$\left \langle {\bf q}_0 \right \rangle $} is calculated by using a homogenized tensor (Eq. (45) in Ref. \cite{Caillerie2012}):
\be\label{q_0_par_G^H}
\Couleur{\left \langle {\bf q}_0 \right \rangle = -\Mat{G}^\mathrm{H}.\nabla_{\bf X} p_0},
\ee
the latter having the form (Eq. (46) in Ref. \cite{Caillerie2012}):
\be\label{G^H}
\Couleur{\Mat{G}^\mathrm{H} = \left \langle \Mat{G}.(1+\nabla_{\bf Y} \chi)  \right \rangle}.
\ee
Note that, with Eq. (\ref{q_0_par_G^H}), the homogenized PDE (\ref{Eq for p-Hom}) has just the same form as the starting ``microscopic" equation (\ref{Eq for p}). However, the homogenized tensor \Couleur{$\Mat{G}^\mathrm{H}$} is not in general the average of the microscopic tensor \Couleur{$\Mat{G}$}: in Eq. (\ref{G^H}), the spacetime vector \Couleur{$\chi$} is the solution of a boundary value problem on the spacetime cell \Couleur{$\Upsilon$} (transposed from Eq. (31) in Ref. \cite{Caillerie2012}). Due to this fact, the application of this method to the actual numerical solution of Eq. (\ref{Eq for p}) looks heavy at first sight: in the relevant quasi-periodic case, we would have to solve that boundary value problem on a representative cell neighbouring an integration point in the macroscopic spacetime domain considered, and this, {\it a priori}, for any integration point --- though possibly less often, depending on the variability that would be found for \Couleur{$\Mat{G}^\mathrm{H}$}.

\section{Conclusion}

In the alternative gravity theory ``SET", electromagnetism in the presence of gravitation demands to introduce an additional energy tensor \Couleur{$\Mat{T}_\mathrm{inter}$}, depending on a scalar field \Couleur{$p$}. This exotic energy tensor might contribute to dark matter. To check this, we developed a model of the interstellar radiation field that provides it as an exact Maxwell field. Then we may calculate the fields \Couleur{${\bf u}$} and \Couleur{$S$} that determine \Couleur{$p$} and \Couleur{$\Mat{T}_\mathrm{inter}$} through the PDE (\ref{dp/dt}). But the very quick variation of \Couleur{${\bf u}$} and \Couleur{$S$}, both in time and in space, makes it unfeasible to integrate (\ref{dp/dt}) on a galactic scale. \\

Therefore, we discussed the application of the homogenization theory. Together with the choice of the precise form for the PDE: either (\ref{Eq for p}) or (\ref{dp/dt}), each choice of a set of variables for the homogenization: time variable, space variables, or spacetime variables, determines a specific homogenization process. 
\bi

\item For Eq. (\ref{dp/dt}) with the time variable alone (``time homogenization"), the PDE stays unchanged but with time-averaged fields: Eq. (\ref{average eq}). However, the data fields that are obtained after the time-averaging, \Couleur{$\bar{{\bf u}}$} and \Couleur{$\bar{S}$}, have still the very rapid {\it spatial} variation, at the scale of the typical wavelength. As a result, it is numerically unfeasible to integrate Eq. (\ref{average eq}) on a galactic scale. 

\item For Eq. (\ref{dp/dt}) with the space variables alone (``spatial homogenization"), an equation for the spatially-averaged unknown field \Couleur{$P=\overline{p_0}$} is obtained if the integral (\ref{Integ_u.grad p_1}) can be neglected, and that equation again is the same as the starting equation, though with spatially-averaged data fields \Couleur{$\overline{{\bf u}}$} and \Couleur{$\overline{S}$}, Eq. (\ref{Integ_order_0-2}). This equation seems to be a tractable one for an integration at the galactic scale --- provided the spatial averaging at a given time also sufficiently smooths out the time variation.

\item Equation (\ref{Eq for p}) for \Couleur{$p$} has precisely the same form as the stationary heat conduction equation for the temperature studied in Ref. \cite{Caillerie2012}, except for the fact that here it is a spacetime equation instead of a spatial equation. Hence, many results of Ref. \cite{Caillerie2012} translate immediately into results for the ``spacetime homogenization" of Eq. (\ref{Eq for p}). The homogenized equation, Eq. (\ref{Eq for p-Hom}) with (\ref{q_0_par_G^H}), has the same form as the starting equation (\ref{Eq for p}), but here the homogenized tensor \Couleur{$\Mat{G}^\mathrm{H}$} is not the average of the microscopic tensor \Couleur{$\Mat{G}$}. The calculation of \Couleur{$\Mat{G}^\mathrm{H}$} involves the solution of a boundary value problem on the spacetime cell. In the relevant quasi-periodic case, we would have to solve that boundary value problem at many different points in the macroscopic spacetime domain considered.
\ei

{\bf Acknowledgement.} The author is grateful to Denis Caillerie for a helpful exchange.

\end{document}